# Constraining Evidence Strength in Bayesian Tonal-State Inference: A Posterior-Continuity Diagnostic

*Chen-Gia Tsai*

Independent Researcher
cgtsai1971@gmail.com

## Abstract

Effective pitch-evidence strength governs the concentration of tonal-state posteriors. Expert local-key annotations cannot calibrate this strength because they do not specify the posterior distribution over the 24 candidate keys. This study constrains evidence strength through tonal continuity, quantified by a measure λ derived from adjacent-window posterior overlap. Passages with λ > 0.50 exhibit relatively strong tonal continuity. At λ = 0.50, carrying forward the preceding tonal belief is predictively equivalent to starting afresh with a uniform, key-neutral prior. Across 26 subcorpora of the Distant Listening Corpus, median λ remains above 0.50 for changes from major keys to IV, V, and vi throughout the examined range of evidence strengths, whereas changes to i, ii, and iii cross this threshold as evidence strength increases. Because these latter three keys share the departed key's tonic or differ from it by a single key-signature accidental, such changes are expected to retain λ > 0.50 in a substantial proportion of cases. An event share *G* therefore assesses how often this expectation holds as tonal selectivity increases in 4-bar windows. The resulting high-*G* region identifies evidence-strength settings that may be appropriate for tonal-state inference. When λ is extended to cross-scale compatibility, the Presto finale of Mozart's Piano Sonata K. 310 shows that, within this region, a 4-bar C-minor interpretation remains compatible with its 16-bar C-major context. The findings establish tonal continuity as a music-theoretical constraint on evidence strength and illustrate how local and prevailing-key interpretations can coexist when pitch evidence is integrated over nested spans.

### 1. Introduction

Temporal integration span shapes computational tonal analysis: shorter windows emphasize local harmonic details, whereas longer windows highlight broader tonal regions (Weiß, Klauk, et al., 2020). This dependence on temporal span poses a challenge for computational key estimation, particularly in distinguishing tonicization from modulation. In music theory, this distinction depends on the duration and structural significance of a temporary tonal center (Aldwell & Schachter, 2003). Applied harmony may briefly favor a secondary tonic while the prevailing key persists over a longer passage. Nápoles López et al. (2020) found that some local-key predictions aligned more closely with tonicization than modulation annotations. Conventional single-chain hidden Markov model (HMM) key trackers regulate tonal persistence partly through key-state transition probabilities (Nápoles López et al., 2019; Noland & Sandler, 2006). These implementations do not separately infer key-state pos-teriors from pitch evidence integrated over different spans.

A further challenge in key estimation arises when converting pitch evidence into probabilities over candidate keys. Key-profile methods assess how well a passage's pitch-class distribution matches major- or minor-key templates (Temperley, 1999, 2007), but deriving posterior probabilities from these fits requires an evidence scale. Expert key annotations cannot calibrate this evidence scale because a single key label does not specify a listener's probability distribution over the 24 candidate keys.

In Western classical music, uncertainty about the prevailing key is a familiar listening experience and sometimes a deliberate compositional effect. In the development section of the first movement of Beethoven's Symphony No. 4, for example, a diminished-seventh harmony is prolonged for 24 measures, delaying the establishment of a clear tonal center (Ferraguto, 2019). Another example is the *bifocal close*, in which a half cadence on the home key's dominant precedes its tonicization as the secondary key (Winter, 1989). At the close, listeners may entertain two key interpretations: the home key, with the harmony functioning as V, and the dominant key, with the same harmony functioning as I. Weiß, Klauk, et al. (2020) identify such a close in Beethoven's Piano Sonata Op. 2 No. 3. Such context-dependent ambiguity, together with differences across temporal integration spans, may contribute to variability in expert key annotations (Ding et al., 2025; Gedizlioğlu & Erol, 2024; Hentschel et al., 2025; Weiß, Schreiber, & Müller, 2020). Evidence strength therefore requires constraints beyond categorical key identity.

Posterior overlap provides a quantitative basis for constraining evidence strength. Tsai (2026) derived $\lambda$ from posterior overlap as a quantitative measure of *tonal continuity* between successive contexts. In Western art music, $\lambda$ rarely fell below 0.50, indicating that tonal continuity generally remained high across successive passages. At $\lambda = 0.50$, carrying forward the preceding tonal belief is predictively equivalent to starting afresh with a uniform, key-neutral prior under equal prior mixture weights. Building on this framework, the present study examines how much tonal selectivity—the degree to which a posterior favors a particular key—can increase while maintaining $\lambda > 0.50$ across key changes expected to retain substantial tonal continuity.

Music theories of tonal relatedness provide a basis for such expectations. Schoenberg (1969) distinguished closely and remotely related tonal regions, while Lerdahl (2001) formalized distances between keys within a hierarchical tonal pitch space. Closely related keys generally share more tonal material (Parncutt, 2024). This suggests that key changes between them preserve greater tonal continuity than those between remote keys. This expectation motivates examining how tonal continuity across different key relations changes as evidence strength increases. The present study assumes that, for key changes of intermediate tonal distance, an appropriate evidence strength allows the median $\lambda$ across such changes to remain slightly above 0.50. Such changes are expected to retain less tonal continuity than changes between very closely related keys (potentially $\lambda > 0.70$), while avoiding the pronounced loss of continuity associated with remote-key changes ($\lambda < 0.50$).

The study first compares median $\lambda$ curves for annotated key changes from major keys across subcorpora of the Distant Listening Corpus (Hentschel et al., 2025), using 4-bar windows. Some key relations are expected to retain substantial tonal continuity, yet their median $\lambda$ falls below 0.50 at sufficiently high evidence strengths. These relations are selected for an event-share analysis.

Tonal relatedness alone, however, does not determine posterior continuity: even changes between the same pair of keys may exhibit different degrees of overlap because their adjoining passages provide different pitch evidence. The event share $G$ therefore assesses how often increasing tonal selectivity can coexist with $\lambda > 0.50$ across individual key changes. The resulting high-$G$ region is taken to represent an appropriate range of evidence-strength settings.

Because $\lambda$ depends only on the overlap of two normalized posteriors, the same measure can also assess tonal compatibility across temporal integration spans by comparing nested windows of different lengths. To examine this cross-scale compatibility, a case study of a passage involving mode mixture in the Presto finale of Mozart's Piano Sonata K. 310 applies the measure within the high-$G$ region identified by the corpus analysis. The analysis examines whether a 4-bar posterior can support a clear minor-key interpretation while remaining compatible with the prevailing major-key interpretation inferred over a longer temporal span. The case study thus addresses the challenge of temporal integration span identified at the outset: different spans may yield distinct tonal interpretations. Throughout, music-theoretical expectations of tonal continuity across key changes and cross-scale compatibility between local and prevailing-key interpretations provide constraints on evidence strength.

# 2. Model and Methods

## 2.1 Tonal-state posterior, evidence strength, and leakage threshold

Let $x$ denote the normalized, duration-weighted pitch-class distribution of a window and $f_i$ the normalized Temperley–Kostka–Payne (TKP) profile for candidate key $i$ (Temperley, 2007). Following Bissiri et al. (2016), the KL profile loss yields likelihood-like support

$$L_i(x; N) = \exp[-N D_{\mathrm{KL}}(x \| f_i)]. \quad (1)$$

With a uniform reference prior over the $J$=24 major and minor keys, the tonal-state posterior is

$$P_i(x; N) = \frac{L_i(x;N)}{\sum_{j=1}^{J} L_j(x;N)}, \qquad J = 24. \quad (2)$$

Effective evidence strength is parameterized for a trailing window of $B$ bars as

$$N = k\,(h_{\mathrm{b}} B)^{\alpha}, \qquad k > 0, \quad \alpha \geq 0. \quad (3)$$

Here $h_{\mathrm{b}}$ is the harmonic density estimated separately for each movement, as defined in Section 2.2. The parameter $k$ is the overall evidence gain, and $\alpha$ is the scaling exponent ($0 \leq \alpha \leq 1$ in the principal grid). The product $h_{\mathrm{b}}B$ approximates the number of annotated harmonic events in a $B$-bar window. $N$ is an effective evidence strength, not a count of independent observations (Tsai, 2026). For $\alpha > 0$, evidence strength increases with $B$. Consequently, identical relative fits to the key profiles yield more concentrated posteriors over longer windows.

For independently inferred tonal-state posteriors $p$ and $q$ from adjacent nonoverlapping windows, their overlap is

$$\langle p, q \rangle = \sum_{i=1}^{J} p_i\, q_i. \quad (4)$$

Tsai (2026) defined the *leakage threshold* from this overlap:

$$\lambda = \frac{J\langle p,q \rangle}{1 + J\langle p,q \rangle}. \quad (5)$$

At $\lambda = 0.50$, the inherited tonal belief and a uniform, key-neutral prior receive equal posterior support when given equal prior weights. A trough in $\lambda$ indicates that the preceding-window posterior provides relatively weak predictive support for the pitch evidence in the following window. Thus, $\lambda$ serves as a quantitative measure of tonal continuity across successive windows.

Defining $S(r) = \sum_i r_i^2$, $S_\Sigma = S(p) + S(q)$ and $D^2 = \|q-p\|_2^2$, the exact identity is

$$\langle p, q \rangle = \frac{S_\Sigma - D^2}{2}, \qquad \lambda = \frac{J(S_\Sigma - D^2)}{2 + J(S_\Sigma - D^2)}. \quad (6)$$

Thus posterior concentration and displacement jointly determine $\lambda$ (Tsai, 2026).

## 2.2 Corpus and preprocessing

This study uses version 3.2 of the Distant Listening Corpus (DLC; Hentschel et al., 2025), which includes corpora annotated by the Digital and Cognitive Musicology Lab (DCML), such as the Annotated Beethoven Corpus (Neuwirth et al., 2018) and Annotated Mozart Sonatas (Hentschel et al., 2021). Movements lacking usable harmony tables were excluded. The applicability of the TKP profiles to each subcorpus was screened on its stable single-key 4-bar windows, defined as trailing windows of bars $t-4$ through $t-1$ at barline $t$ in which a single absolute DCML key holds throughout and no annotated key boundary (including its within-bar onset) occurs between $t-8$ and $t+4$. Subcorpora were retained only if they provided at least 100 such windows and the annotated key ranked among the three TKP profiles with the lowest KL losses in more than 70% of them. This screening checks that the profiles rank the annotated key highly in stable passages. The same windows provide the reference distribution for the out-of-key limit of Section 2.3. Peri's Euridice, Monteverdi's madrigals, and Frescobaldi's Fiori musicali were excluded separately because their harmonic practices do not consistently conform to the common-practice major–minor tonal framework assumed in this study.

Each measure-count (*mc*) entry in the measures table is treated as one analytical bar. Notes extending across bar boundaries are apportioned according to their sounding duration within each bar, and durations are aggregated by pitch class (MIDI pitch modulo 12). A window is eligible only if its bars form a continuous stretch of the notated timeline and contain at least 0.5 quarter notes of sounding duration. Movement-level harmonic density, $h_{\mathrm{b}}$, is calculated as the number of harmony annotations with a nonempty numeral field divided by the number of analytical bars.

## 2.3 Key-change events and relation-level tonal continuity

A relation-level screening was conducted to identify key changes that could provide informative constraints on evidence strength. The analysis considered annotated changes from major keys to other major or minor keys. Music-theoretical expectations suggest that closely related keys should retain substantial tonal continuity, whereas more remote keys may exhibit weaker continuity. Relations of intermediate tonal relatedness are therefore expected to be particularly informative: their posterior overlap may remain above $\lambda = 0.50$ at moderate evidence strengths but fall below this threshold as the posteriors become more selective.

Three relations were expected to exhibit this behavior: changes from a major key to its parallel minor (I→i), to the minor key a whole step above (I→ii), and to the minor key a major third above (I→iii). The parallel minor shares the tonic of the departed key, while ii and iii differ from it by only one accidental in the key signature. Taking C major as the reference, these three relations lead to C minor (i), D minor (ii), and E minor (iii), respectively. Their symmetrized TKP-profile KL divergences (summed over both directions) from C major are 0.88, 0.81, and 0.61 nats, respectively. By comparison, the three most closely related keys—G major (V), F major (IV), and A minor (vi)—have smaller divergences of 0.45, 0.45, and 0.37 nats, respectively. Thus, I→i, I→ii, and I→iii are expected to retain substantial shared tonal support while allowing increasing evidence strength to differentiate the two keys.

For each candidate key change, both the departed and entered local-key regions were required to last at least $B$ bars, allowing complete, contiguous windows on either side of the boundary. This duration requirement served as an inclusion criterion and did not imply that the new key was established at the level of musical form. A change in the first half of a bar was aligned to that bar, whereas a later change was aligned to the next. The pre-change window comprised the $B$ bars preceding this boundary, and the post-change window the $B$ bars beginning at it. Both windows used the effective evidence strength $N$ calculated for their shared movement and window length.

To limit the influence of extended chromatic preparation, neither window could contain more than 20% of its sounding duration on pitch classes outside its annotated key. Otherwise, posterior overlap might reflect shared chromatic content rather than the tonal relation between the departed and entered keys. For major keys, the in-key collection comprises the diatonic scale. For minor keys, it includes the natural minor scale and the raised sixth and seventh degrees, so both forms of these degrees count as in key (nine pitch classes). The pre-change window was evaluated against the departed key and the post-change window against the entered key. The 20% limit was set near the 90th percentile of out-of-key duration in the stable single-key windows described in Section 2.2, allowing chromatic content within the upper range observed in stable key regions.

For the initial relation-level screening, median λ was computed as a function of $k$ for each key relation represented by at least 30 qualifying changes, using 4-bar windows at $\alpha = 0.30$. Relations were selected for the event-share analysis when their median λ crossed the continuity criterion of 0.50 as evidence strength increased, despite a music-theoretical expectation of substantial tonal continuity. Relations that retained $\lambda > 0.50$ throughout the examined range offered little discrimination among evidence strengths, while relations exhibiting much weaker continuity were less suitable for this constraint.

For each event $e$, $\lambda_e(N)$ was evaluated at 1,200 logarithmically spaced N values from 0.01 to 1,000, with linear interpolation in ln $N$. For the three selected relations, the critical k at which median λ first fell to 0.50 after its maximum was obtained by linear interpolation at $k$ steps of 0.05. Its 95% interval was estimated from 1,000 movement-cluster bootstrap replicates within each relation, with joint movement resampling for pairwise comparisons.

## 2.4 Event share

For each event, $N_{\text{peak},e}$ denotes the evidence strength at maximum overlap and $N_{\text{cross},e}(\lambda^*)$ the first subsequent descent to a continuity criterion $\lambda^*$. The principal criterion $\lambda^* = 0.50$ is the point of equal support for inherited belief and the key-neutral alternative. If no crossing is observed by $N = 1{,}000$, the computational upper limit $N_{\text{cross},e} = 1{,}000$ is used, and events whose maximal λ does not exceed $\lambda^*$ have no qualifying interval. A setting contributes to the event share $G$ only when the movement-specific evidence strength of event $e$ lies between $N_{\text{peak},e}$ and $N_{\text{cross},e}(\lambda^*)$:

$$G(k, \alpha; B, \lambda^*) = \frac{1}{E}\sum_{e=1}^{E} 1\left[ N_{\text{peak},e} < k\left(h_{\text{b},e}B\right)^{\alpha} < N_{\text{cross},e}(\lambda^*) \right] \quad (7)$$

where $E$ is the total number of events satisfying the event-selection criteria, including those with no qualifying interval, which contribute zero to $G$. In Equation 7, $h_{\text{b},e}$ denotes the movement-level harmonic density $h_\text{b}$ for the movement containing event $e$. This operational definition excludes the rising branch, on which increasing posterior concentration still increases overlap.

The principal analysis evaluates $G$ over $1 \le k \le 8$ and $0 \le \alpha \le 1$, using 4-bar windows and a continuity criterion of $\lambda^* = 0.50$. A high-$G$ region is defined as the set of settings with $G > G^*$, with $G^*$ varied from 0.40 to 0.48 in steps of 0.01. Two robustness analyses examine the effects of relaxing the continuity criterion to $\lambda^* = 0.45$ with 4-bar windows and extending the window length to 8 bars while retaining $\lambda^* = 0.50$. The 8-bar analysis requires both the pre-change and post-change annotated key regions to last at least 8 bars and applies the same 20% out-of-key duration limit to each window. The setting $(\alpha, k) = (0.30, 3.5)$, adopted from Tsai (2026), serves as a fixed reference for comparing the three configurations and for the case study.

### 2.5 Multiscale case study

A case study examines how the posterior behaves across window lengths within a single movement and how it relates to the annotated local key. For the Presto finale of Mozart's Piano Sonata in A minor, K. 310, the posterior concentration $S$ of the trailing window $[t - B, t)$ is computed at every barline $t$ for $B = 2, 4, 8$, and 16, together with the trailing λ between the two most recent windows, $[t - 2B, t - B)$ and $[t - B, t)$. All four scales use the reference setting $(\alpha, k) = (0.30, 3.5)$, so that the evidence strength of each scale follows from $N = k(h_\flat B)^\alpha$ alone. The maximum a posteriori (MAP) key and the DCML local-key annotation are compared for each window. The analysis covers mm. 15–44, which span the annotated change from A minor to C major and the first phrases of the second key area.

Equation 5 is also applied across scales. Whereas λ compares adjacent windows of the same length, the cross-scale tonal compatibility $\lambda_{cross}$ compares the posteriors of a short and a long window that end at the same barline. $\lambda_{cross}$ is evaluated over $1 \le k \le 8$ at $\alpha = 0.30$. In addition, the range of $\lambda_{cross}$ over all grid settings $(\alpha, k)$ within the high-$G$ region is obtained for each $G^*$ of the principal analysis. Cross-scale tonal compatibility is examined as a graded quantity, with $\lambda_{cross} = 0.50$ serving as the equal-support criterion. Because both posteriors are derived under the same uniform reference prior, Equation 5 retains this interpretation across scales: at $\lambda_{cross} = 0.50$, using the short-span posterior to predict the long-span evidence is formally equivalent to using a uniform, key-neutral prior, and vice versa.

ChatGPT (OpenAI) and Claude (Anthropic) were used to assist with discussion and refinement of model formulations and analytical strategies, development and debugging of analysis code, and manuscript language editing. The author reviewed the AI-assisted work and takes full responsibility for the analyses, interpretations, and content of the manuscript.

## 3. Results

### 3.1 Modulation relations and the selection of diagnostic events

Of the 40 subcorpora, seven fall short of the top-three criterion of the applicability screening (Schulhoff, Bartók, Sweelinck, Ravel, Debussy, Schütz's Kleine geistliche Konzerte, and Liszt), four supply fewer than 100 stable 4-bar windows (Handel, F. Couperin's harpsichord pieces, Schumann's Kinderszenen, and Poulenc), and three are excluded on stylistic grounds, leaving 26 subcorpora.

At $\alpha = 0.30$, eight relations from a major key meet the frequency requirement (Figure 1). The remaining relations, including I→II and I→♭III, are represented by fewer than 30 qualifying changes and are not shown. Changes to IV, V, and vi keep median λ above 0.50 throughout $k = 1$–8. Changes to iii, ii, and i fall below it at successively smaller $k$. Changes to v and ♭VI show still less tonal continuity. Because changes to IV, V, and vi retain continuity across the whole range, they cannot discriminate among evidence strengths, whereas I→i, I→ii, and I→iii cross 0.50 within it and therefore supply the events for the event-share diagnostic. Their critical $k$ is 3.95 for I→i, 4.66 for I→ii, and 5.21 for I→iii, although the bootstrap intervals of their pairwise differences all contain zero. The display reference $k = 3.5$ lies below all three critical values.

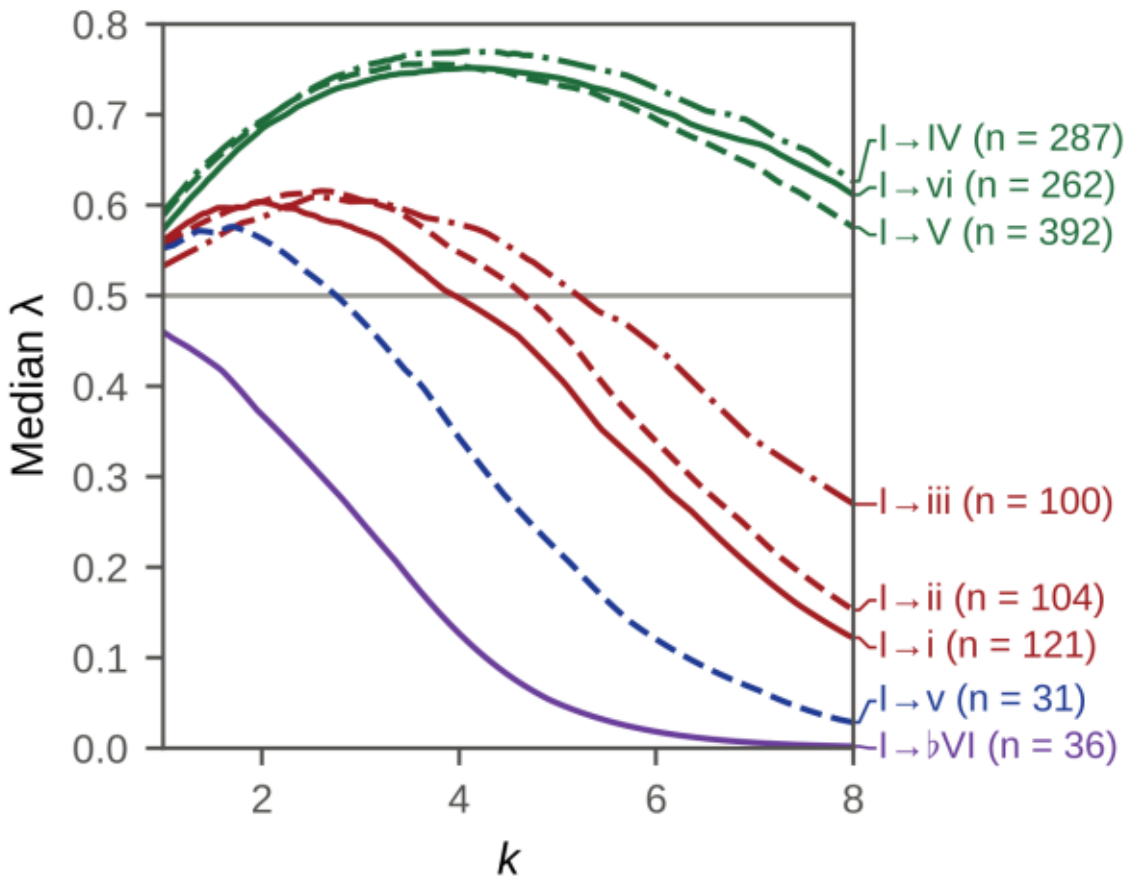

**Figure 1.** Median λ versus $k$ at $\alpha = 0.30$ for the eight relations from a major key represented by at least 30 key changes in the 26 retained subcorpora ($B = 4$; both regions at least 4 bars long; at most 20% out-of-key duration per window). Labels indicate the number of qualifying changes, and the gray horizontal line marks $\lambda = 0.50$.

### 3.2 Event share across continuity criteria and window lengths

Of the 367 I→i, I→ii, and I→iii key changes meeting the region-length and window requirements, 42 exceed the out-of-key limit, 35 through the pre-change window alone. The 90th percentile of out-of-key duration in stable reference windows is 0.22 in major and 0.17 in minor. The resulting 325 key changes comprise 121 I→i, 104 I→ii, and 100 I→iii events from 253 movements in 24 retained subcorpora. Of these, 310 attain a maximum λ above 0.50, whereas 15 do not and have no qualifying interval. Among the 310, the median $N_{peak,e}$ is 4.55 and the median $N_{cross,e}(0.50)$ is 8.79. Nine of them show no crossing by the computational upper limit of $N$ = 1,000 and enter the latter median at that limit. In these nine events the pre- and post-change windows favor the same best-fitting key, so their high continuity does not indicate that the model distinguishes the annotated key change. With 8-bar windows, the same requirements yield 191 events (82 I→i, 50 I→ii, and 59 I→iii) from 160 movements in 21 subcorpora after 25 events are excluded by the out-of-key limit. Of these, 178 attain a maximum λ above 0.50.

With 4-bar windows and $\lambda^*$ = 0.50 (Figure 2a), $G$ forms a single oblique ridge in the (α, $k$) plane, following the trade-off between gain and density scaling that leaves $N$ approximately constant. $G$ peaks at 0.489 at (0.25, 3.65) and equals 0.449 at the display reference (0.30, 3.5), which lies within the ridge. Because $G$ at α = 0 is already close to its maximum, continuity alone does not identify the density-scaling exponent, and the ridge constrains the effective evidence strength $N$ rather than $k$ and α separately. At $G^*$ = 0.43, the high-$G$ region spans approximately 2.95 ≤ $k$ ≤ 3.55 at α = 0.30.

Relaxing the continuity criterion to $\lambda^*$ = 0.45 (Figure 2b) extends the qualifying intervals: all 325 events rise above this criterion, $G$ increases across the plane to a maximum of 0.585, and the ridge broadens. At α = 0.30, the ridge maximum shifts only from $k$ = 3.25 to 3.20, and $G$ at the reference setting is 0.545.

With 8-bar windows (Figure 2c), the 191 events yield a ridge of the same orientation, with a maximum of 0.550 and $G$ = 0.482 at the reference setting, where 92 events lie on their qualifying branch and 66 have crossed below λ = 0.50. At α = 0.30, the ridge lies at $k$ = 3.05, slightly below its 4-bar position. The constraint identified with 4-bar windows is approximately preserved at the longer integration span.

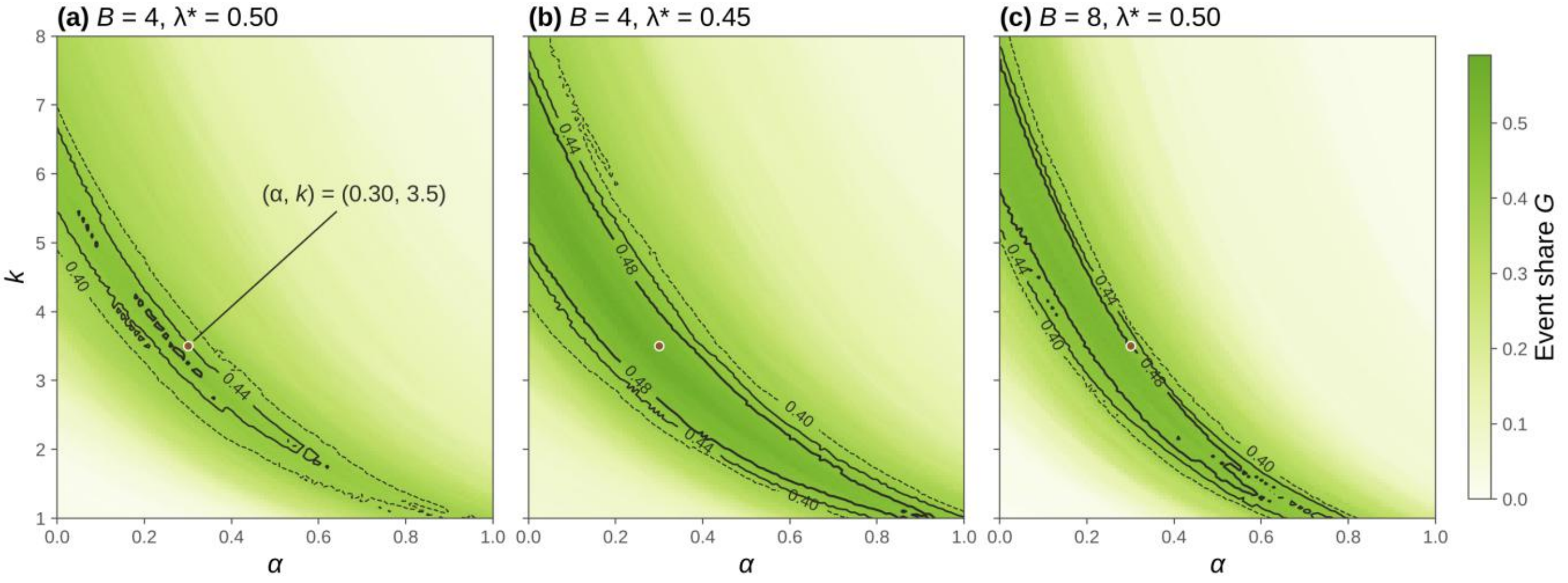

**Figure 2.** Event share $G$ over the (α, $k$) plane for (a) 4-bar windows with continuity criterion $\lambda^*$ = 0.50, (b) 4-bar windows with $\lambda^*$ = 0.45, and (c) 8-bar windows with $\lambda^*$ = 0.50. Shading gives the share of the annotated I→i, I→ii, and I→iii key changes (325 events in panels a and b; 191 in panel c) that lie beyond the maximum of their own λ curve while retaining $\lambda > \lambda^*$, on a scale common to the three panels; labeled contours mark $G$ = 0.40 (dashed), 0.44, and 0.48. Brown dots mark the display reference (α, $k$) = (0.30, 3.5).

## 3.3 Cross-scale compatibility in K. 310

Figure 3a shows scale-dependent tonal interpretations in the Presto finale of Mozart's Piano Sonata K. 310, using expert local-key and harmonic annotations from the Annotated Mozart Sonatas corpus (Hentschel et al., 2021), included in the Distant Listening Corpus (Hentschel et al., 2025). The annotated local key changes from A minor to C major at m. 21 and remains C major thereafter. The 16-bar posterior follows a similar course, passing from a concentrated A-minor state through a trough of low concentration to a C-major peak near m. 40. Its transition lags behind the annotated key change because each trailing window integrates pitch evidence up to its endpoint.

Within the annotated C-major region, mm. 29–31 prolong C-minor tonic harmony (i). The progression then moves through $iv^6$ to a half cadence on V in m. 32. During this phrase, the 2- and 4-bar posteriors favor C minor. The 2-bar posterior remains relatively diffuse and changes its MAP key more frequently, reflecting the limited pitch evidence available within this shorter window. The 4-bar posterior achieves greater concentration, with its C-minor maximum exceeding the neighboring C-major maxima.

The C-minor episode is not annotated as a key change and therefore lies outside the 325 events used to compute $G$. The adjacent-window $\lambda$ at the 4-bar scale also falls just below 0.50. The phrase recalls the head motive of the movement's opening theme, preserving thematic continuity despite its local tonal reorientation. At the end of m. 32, the 4-bar posterior (mm. 29–32) favors C minor, whereas the 16-bar posterior (mm. 17–32) favors C major, with $\lambda_{cross} = 0.50$ at $k = 3.5$ (Figure 3b). Across the high-G interval of $2.95 \leq k \leq 3.55$ at $\alpha = 0.30$ (shaded in Figure 3b), the 4-bar C-minor posterior probability rises from 0.53 to 0.65, while $\lambda_{cross}$ remains near the equal-support criterion, declining only from 0.55 to 0.49. Table 1 extends this comparison to all settings within the high-$G$ region for each $G^*$. As $G^*$ increases, the minimum of $\lambda_{cross}$ rises and exceeds 0.50 for $G^* \geq 0.46$. Thus, the local minor-key interpretation gains posterior support without a corresponding collapse in cross-scale compatibility. At lower $k$, $\lambda_{cross}$ is slightly higher, but the 4-bar posterior assigns C minor little probability. At higher $k$, the 4-bar posterior commits to C minor and $\lambda_{cross}$ falls toward zero as the two scales concentrate on different keys. Only an intermediate range of evidence strengths allows the mode-mixture phrase to retain its minor-mode character while remaining compatible with its longer-span C-major context.

| $G^*$ | Grid settings | $k$ ($\alpha = 0.30$) | $\lambda_{cross}$ |
|---|---|---|---|
| 0.40 | 1,745 | 2.65–3.90 | 0.42–0.59 |
| 0.41 | 1,510 | 2.65–3.85 | 0.44–0.59 |
| 0.42 | 1,260 | 2.75–3.75 | 0.44–0.59 |
| 0.43 | 963 | 2.95–3.55 | 0.47–0.58 |
| 0.44 | 668 | 3.00–3.50 | 0.48–0.58 |
| 0.45 | 451 | 3.15–3.45 | 0.49–0.57 |
| 0.46 | 304 | 3.20–3.40 | 0.50–0.57 |
| 0.47 | 167 | 3.20–3.30 | 0.51–0.57 |
| 0.48 | 24 | 3.25 | 0.52–0.55 |

**Table 1.** Cross-scale tonal compatibility $\lambda_{cross}$ at the end of m. 32 of K. 310/iii over the high-$G$ region for each threshold $G^*$. Grid settings gives the number of ($\alpha$, $k$) settings with $G > G^*$ in Figure 2a, and the third column gives the range of $k$ in this region at $\alpha = 0.30$. Throughout, the 4-bar posterior favors C minor and the 16-bar posterior favors C major.

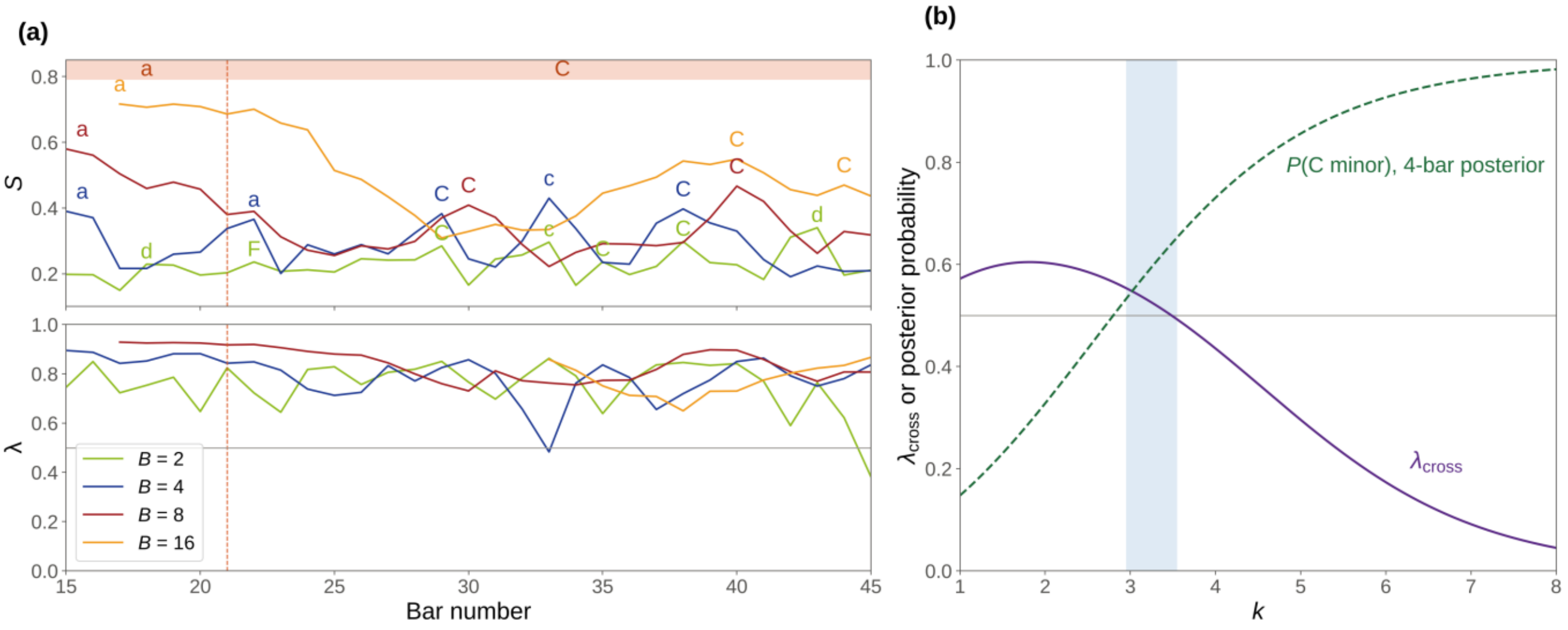

**Figure 3.** Multiscale posteriors in Mozart, Piano Sonata K. 310, third movement. **(a)** mm. 15–44 at ($\alpha$, $k$) = (0.30, 3.5). Upper: posterior concentration $S = \Sigma_i p_i^2$ of the trailing window $[t - B, t)$ at each barline $t$ for $B = 2$, 4, 8, and 16; local maxima are labeled with the MAP key, and the band at the top shows the DCML local-key annotation (A minor, then C major from m. 21, dashed line). Lower: trailing $\lambda$ between the windows $[t - 2B, t - B)$ and $[t - B, t)$. **(b)** Cross-scale tonal compatibility $\lambda_{cross}$ at the end of m. 32 as a function of $k$ ($\alpha = 0.30$), between the 4-bar posterior (mm. 29–32) and the 16-bar posterior (mm. 17–32) (solid line), and the probability of C minor in the 4-bar posterior (dashed line). The shaded band marks the range of $k$ (2.95–3.55) in which the 4-bar event share exceeds $G^* = 0.43$ at $\alpha = 0.30$ (Figure 2a). Gray horizontal lines mark 0.50.

## 4. Discussion

Converting key-profile fits into posterior probabilities requires an evidence scale, which categorical key annotations cannot supply: a label does not specify how support is distributed over the 24 candidate keys. The present results show that tonal continuity across annotated key changes provides such a constraint. For changes from a major key to its parallel minor (i), to ii, and to iii, the share of events whose posterior overlap has passed its maximum while retaining $\lambda > 0.50$ forms an oblique ridge in the (α, $k$) plane. This ridge persists when the continuity criterion is relaxed to $\lambda > 0.45$ and when the window length is extended from 4 to 8 bars while retaining $\lambda > 0.50$. At higher evidence strengths beyond the ridge, posteriors sharpen toward different keys and lose the tonal support shared by adjacent contexts. In the Mozart case study, an evidence strength within the ridge likewise keeps a 4-bar C-minor interpretation compatible with its 16-bar C-major context. Across successive windows in the corpus and nested windows in the Mozart passage, the high-$G$ region identifies evidence strengths at which a posterior can favor a key while retaining substantial tonal support shared with related contexts.

## 4.1 Continuity-based constraints on evidence strength

In this analysis, expert annotations locate key changes and characterize their contexts, including the subcorpus screening, but they do not serve as targets for posterior concentration. The constraint on evidence strength comes from the posteriors themselves. Its music-theoretical content lies in the modulation relations. Thompson and Cuddy (1989, 1992) demonstrated that perceived key movement is sensitive to modulation distance and direction, providing perceptual grounds for examining tonal continuity separately across key relations. In the present analysis, changes to IV, V, and vi keep λ above 0.50 throughout the examined range and therefore cannot discriminate among evidence strengths.

Changes to i, ii, and iii supply that discrimination. These keys share the departed key's tonic or differ from it by a single accidental, so substantial shared tonal support is musically expected. Nevertheless, their posterior overlap falls below the equal-support criterion on the descending branch as the posteriors sharpen (Figure 1). These relations therefore show how much sharpening the expected support can withstand. Some loss of continuity is expected even for these relations, particularly when a key change lacks harmonic preparation (Section 4.3). The relaxed criterion $\lambda^* = 0.45$ therefore provides a robustness check: the ridge broadens but remains at a similar position, indicating that the constraint does not depend critically on the exact value of the equal-support criterion.

Notably, the high-$G$ ridge also aligns with the admissible evidence-strength region derived by Tsai (2026) from parallel-mode changes in two Mozart movements with contrasting $h_b$. Settings within the earlier region tend to yield higher event shares than those outside it. Using $G$ to distinguish settings inside and outside the earlier admissible region yields an area under the receiver operating characteristic curve (AUROC) of 0.97 for 4-bar windows and 0.96 for 8-bar windows. Thus, constraints derived from corpus-wide event shares and individual passages converge on similar evidence-strength settings.

## 4.2 Tonal persistence and hierarchical tonal representation

The relation-dependent pattern also bears on how key-finding models represent tonal persistence. Conventional single-chain HMM key trackers regulate persistence through key-state transition probabilities: high self-transition probabilities suppress frequent changes when short-span pitch evidence favors different keys in succession, and some trackers derive transition probabilities from measures of tonal relatedness (Noland & Sandler, 2006). Such matrices can encode a proximity ordering similar to that observed in Figure 1. In the present analysis, this ordering emerges from independently inferred tonal-state posteriors.

The Mozart case study offers a further perspective on transition probabilities and tonal persistence: longer-span pitch evidence can sustain a prevailing-key interpretation even when shorter-span evidence favors another key. In K. 310 (Figure 3), the 4-bar posterior favors C minor, reflecting the pitch evidence of the mode-mixture phrase, while the 16-bar posterior retains the broader C-major context. Across the high-$G$ region at $G^* = 0.43$, $\lambda_{cross}$ ranges from 0.47 to 0.58, indicating that the two key interpretations can coexist while largely preserving cross-scale tonal compatibility.

Hierarchical predictive processing offers a framework for interpreting this coexistence. Slowly evolving higher-level states provide context for faster lower-level representations. Descending predictions can explain away lower-level prediction errors, while residual errors, weighted by their precision, contribute to updating higher-level beliefs (Friston, 2005; Kiebel et al., 2008; Koelsch et al., 2019; Vuust et al., 2022). In the Mozart passage, a higher-level C-major representation could provide a context in which the local C-minor departure is interpreted as mode mixture. Such contextualization might explain away some of the associated lower-level prediction errors without requiring revision of the prevailing-key interpretation. From this perspective, maintaining a longer-span tonal representation may help listeners integrate local deviations into a coherent tonal context. Whether such top-down explanation occurs during music listening remains to be investigated.

While the present framework characterizes tonal interpretations across temporal spans using key-state posteriors, theories of harmonic syntax address a comple-mentary aspect of tonal hierarchy. Rohrmeier's (2011, 2020) models formalize hierarchical preparation and prolongational dependencies among harmonic events. These dependencies convey information about harmonic function and temporal organization that cannot be recovered from aggregated pitch-class distributions alone. By focusing primarily on spans of four bars or longer, the present framework may provide a tonal context for interpreting chord functions at shorter spans. At shorter spans of one or two bars, in turn, harmonic dependencies may be particularly important, as pitch evidence alone may provide insufficient support for a stable key interpretation.

### 4.3 Limitations and future directions

The constraint identified here is descriptive. Even changes between closely related keys do not always retain $\lambda > 0.50$. For example, a change between keys whose signatures differ by a single accidental may fall below this threshold when it lacks harmonic preparation of the kind provided by a pivot-chord modulation. Posterior continuity therefore depends on both the tonal relationship between the keys and the actual pitch evidence in the adjoining windows. Accordingly, the high-$G$ region identifies only the evidence-strength settings at which a relatively large share of the selected key changes retain $\lambda > 0.50$ as their posteriors become more selective. Whether the $\lambda = 0.50$ threshold is perceptually meaningful in individual musical passages remains to be established. Listening experiments could present key-change passages with $\lambda$ values above and below 0.50, using 4-bar windows on either side of each change. Listeners would judge perceived tonal continuity and the extent to which the preceding key remains perceptually relevant. Accounting for key relation, harmonic preparation, and other boundary cues would help determine whether $\lambda$ predicts perceived tonal continuity and whether 0.50 marks a meaningful perceptual distinction. Such findings could provide behavioral constraints on evidence strength.

The multiscale case study is limited to a single passage in Mozart's K. 310, and its cross-scale result remains illustrative. Further corpus analyses could examine whether similar patterns of cross-scale compatibility occur in other contexts involving mode mixture or tonicization. Listening experiments could manipulate the longer-span tonal context while holding the local passage constant, obtaining perceptual measures of the compatibility between local and prevailing-key interpretations. Such measures could provide additional constraints on evidence strength across temporal spans, including the shorter spans relevant to tonicization.

## 5. Conclusion

This study establishes tonal continuity as a music-theoretical constraint on effective pitch-evidence strength in Bayesian tonal-state inference. For annotated I→i, I→ii, and I→iii key changes, the event share $G$ forms an oblique ridge in the ($\alpha$, $k$) parameter space, identifying a range of appropriate evidence-strength settings. In the Presto finale of Mozart's K. 310, an evidence strength within this region allows the short-span and longer-span posteriors to maintain relatively clear C-minor and C-major interpretations, respectively, while remaining compatible across scales. These findings show how music-theoretical expectations of tonal continuity can constrain evidence strength and provide a basis for multiscale probabilistic representations of tonality.

#### Reproducibility

Data availability. The score, harmony annotations, and metadata are available from the Distant Listening Corpus, version 3.2, https://dcmlab.github.io/distant_listening_corpus/ (Hentschel et al., 2025). Repository commits are recorded in the analysis materials.

Code availability. The analysis code and intermediate data are archived at Zenodo (https://doi.org/10.5281/zenodo.24027370).

#### Competing Interests

The author has no competing interests to declare.